# Automatic Attack Script Generation: a MDA Approach

Quentin Goux[1][0000-0002-9924-1837] and Nadira Lammari[2][0000-0002-2966-5300]

[1,2] CEDRIC, Conservatoire National des Arts et Métiers (CNAM), Paris, France
[1]`quentin.goux2.auditeur@lecnam.net`
[2]`ilham.lammari@lecnam.net`

**Abstract.** It is widely recognized that practical exercises are crucial for teaching cybersecurity in higher education. However, their setup is not only expensive, time-consuming, and prone to numerous errors, but also requires technical and programming skills to create attack contexts and scripts. To mitigate these drawbacks, this research work proposes an approach that automatically generates scripts and attack contexts based on informal attack scenario descriptions. To isolate business concerns from technological issues, our approach is aligned with the MDA development method. A formal language is proposed to express our Computation Independent model. We rely on the TOSCA standard to describe our Platform Independent Model. We also allow through our approach the generation of several Platform Specific Models. Hence, this research work contributes not only to the overall improvement of attack implementations for cybersecurity training but also to their reuse on various platforms.

**Keywords:** attack scenario modeling, attack context modeling, MDA approach, script generation.

## 1 Introduction

The limited knowledge or awareness of risks and security procedures across different types of internet users or employees, coupled with the new era of connectivity, provides fertile ground for adversaries to exploit security breaches and perform attacks that cause significant damage to IT assets and end-users. By providing cybersecurity training opportunities, raising awareness, and improving educational offering, we can contribute to overcoming this situation. Indeed, while raising awareness among beginners allows them to learn how to better protect themselves or their assets, advanced training provides trainees with cybersecurity skills through practical courses with realistic scenarios depicting situations that occur. In this regard, cyber training environments (e.g. cyber ranges) supply a controlled and isolated infrastructure (machines, networks, tools, etc.) that facilitate the implementation of cyber training exercises.

However, the process of preparing and configuring a cyber training exercise is usually done manually, which can be time consuming, error prone, and entails a lot of effort and advanced skills. Specifically, the training sessions preparation involves surveying recent attacks and vulnerabilities that are documented by a variety of sources. The envisioned attacks could be expressed and modeled in different ways [1, 2]. Moreover, the translation in a cyber environment is not straightforward because there is no unified model. A significant effort is also devoted to setting up the cyber training platform, which requires configuring and deploying the IT assets. Furthermore, despite the considerable effort required for a cyber training, it quickly becomes obsolete as it cannot be easily adapted to the rapid evolution of attacks and IT technologies.

Motivated by the above, this research work aims to improve the efficiency of the cybersecurity exercise production process by reducing efforts and error prone tasks, both during the setup phase and during the execution phase. It proposes an approach for the automatic generation of attack scripts from formal descriptions of their associated scenarios.

Given the syntactic and semantic heterogeneity of attack models and the diversity of knowledge they incorporate, we rely on a unified attack model that we introduced in [3]. Our unified attack model formalizes the description of the attack scenarios to be used in training exercises. This formalization is based on an attack scenario description language that we have also defined. In this same publication we have also provided an attack context model for the description of all IT resources involved in an attack.

Our approach adopts the MDA development approach. The latter promotes rapid code generation. It describes an application through three successive abstraction levels: computation independent, platform independent end platform specific levels. A model is associated with each level of abstraction. In our approach the unified attack model is used as a CIM (Computation Independent Model) to formally express the attack requirements. For its instantiation, a user interface is provided. Thereafter, our approach considers the attack requirements and provides, through a series of automatic transformations, an implementation of the attack script and context in an ad hoc execution platform.

The rest of this paper is organized as follows. Section 2 reports on related work. Section 3 provides an overview of our proposed approach, which is also illustrated through an example of an attack scenario. Sections 4 to 6 are dedicated to each phase of our approach from the requirement specification to the implementation. The last section concludes and suggests potential avenues for future research.

## 2      Related work

In the literature, graph-based representation of attacks was introduced as a modeling means for several automations. The most popular ones are attack trees and attack graphs [4]. On the side of attack trees, [5] presented threat trees as a logical structures describing an iterative decomposition, from the primary threat objective to individual steps by going through intermediate goals. Then [6] broadcasted attack trees by generalizing the formal tree structure to any decomposition of goals into sub-goals, for which effort of (re)formalization were proposed [7]. [8] proposed a way of documenting attacks which enables an organization to establish and reuse relationships between detailed attack patterns through its forest database of attack trees.

On the side of attack graphs, the structure is deduced during the generation. Along the recognition that model information is conditioned by its generation method, [9] presented two opposite algorithmic initiatives, both leveraging data about and underlying network : the backward one, suitable for retrieving attack paths, and the forward one, suitable for exploring possibilities [10, 11].

The literature on attack automation based on those attack models, to the best of our knowledge, remains limited to simulations which stay enshrined in the realm of models. The script generation presented in [12] is specific to their simulation platform, as it requires activity templates ahead of the [11] graph generation from which the scenario shall be extracted. The scripts generated from [13] are specific to the systems playing target for evaluating their security by testing.

Languages constitute another track for the modeling of attacks. In this regard, [14] proposed a distribution of all attack languages in six classes, each gathering some shared properties around for addressing a specific scope. However, no strain is attached to the 'exploit language' class for which there is no broadly accepted standard, and their scripting of attacks is left to general-purpose programming languages. To tackle the challenge of describing from the attacker's point of view, there is a part of the ADeLe language [15] that is dedicated to the exploit. It fosters a causal approach through its 3 sections: the 'precondition', the 'attack' and the 'postcondition'. Once again, the sections either lack formalization or stay open to any general-purpose language.

More recently, [16–18] proposed the Meta Attack Language (MAL) as a Domain Specific Language (DSL) factory for the specification of attack enabling their simulation. DSL instances are built by formalizing attack steps on entities gathered in classes, represented by assets recognized inside the domain. Then, specific analyses, such as the computation of the global time to compromise, may be automated through MAL simulation paths.

Besides education and training, automation of attacks may also benefit cybersecurity evaluations fields, especially pentesting. The approach proposed in [19] states that any automated pentesting solution comprises two main features: attack planning and attack automation, referring the ability to perform the testing process. It also puts the realism of the evaluation environment based on simulation into perspective, with respect to a realistic virtual network. This perspective has also been identified as a concern for learning performance in [20], although it is focused on machine learning, by opposing simulation to emulation.

In the context of cyber ranges [21, 22], which may be simulation or emulation based, stakeholders and roles are grouped and assigned to teams identified by colors. The primary ones are red, for attackers, and blue, for defenders, and white, for instructors [23]. In addition, other teams have been defined to refer to specific points of view [24]. Thus, our approach leans toward the orange team point of view at design phase by fully automating the red team comportment.

## 3      Overview of our script generation approach

Availability of attack scripts is a major requirement for setting up and managing cyber training exercises. When not available, designers of the cyber training exercises implement them based on informal descriptions of attacks. For this purpose, they mobilize their practical know-hows which are the result of their experiences. This unfortunately repetitive task easily becomes tedious. Besides their cybersecurity skills, designers must also mobilize various additional skills: technical, programming and modeling skills. Modeling skills help them to have an initial abstraction of the informal description of the operating mode. As defined in [25]: "*Abstraction is a cognitive means by which engineers, mathematicians and others deal with complexity. It covers both aspects of removing detail as well as the identification of generalizations or common features*". It corresponds, in our case, to an attack model. However, faced with the multiplicity of attack models, designers choose to express the attack scenarios with the model with which they are most familiar, which very often raises the problem of attack scenario reuse.

To allow designers to focus on modeling the attack scenario while abstracting away the technical aspects that are covered by an executable script, we propose, through this paper, an automation of the script generation process. We therefore propose an approach that has an attack scenario as input and returns a script as output.

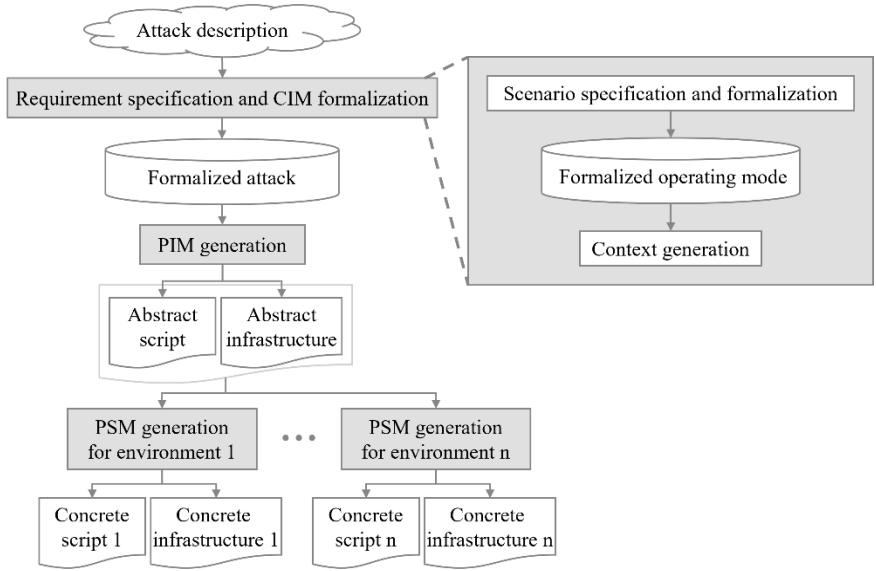

**Fig. 1.** Script generation approach from an attack description

Our approach is a model-based one. It benefits from the advantages of Model Driven Engineering (MDE): rapid code generation, reduction of development errors and consistency maintenance between design and code and finally a reduction of the designer effort. Furthermore, to isolate business concerns from technological issues, it aligns with the Model Driven Architecture (MDA) development approach [26].

MDA is based on the principle of using modelling languages like UML to specify a system at various and successive abstraction levels: the level where its requirements are formulated in terms of how the system will be used in the environment; the one describing the system architecture in a technology-neutral manner and the one specifying how the model is to be implemented using a specific platform. Hence, it enables the generation of multi-platform applications. MDA, as a variant of MDE, associates a model to each level, respectively called Compute Independent Model (CIM), Platform Independent Model (PIM) and Platform Specific Model (PSM). It also recommends the implementation of mechanisms for transforming source models into target models.

Therefore, our approach for the generation of attack scripts encompasses three phases (see Fig. 1). The first one is dedicated to specifying requirements and formalizing the attack context for the attack scenario, for which the script is to be generated. The attack context includes all the resources involved in the attack, mainly those targeted and those mobilized for achieving the attack goal. Thus, the phase corresponds to specifying the CIM model. The CIM model comprises the models of the attack operating mode and context that have been described in [3] and briefly recalled in Section 4. In order to help the user in this specification and ensure conformity to the model, we developed a user interface for instantiating a knowledge base with the scenario description. As shown in the upper right part of Fig. 1, the context is automatically generated from the formalized operating mode. Its model corresponds to the description of the different states it goes through when the scenario unfolds.

The second phase (PIM generation in Fig. 1), aims to generate both the abstract attack script and the abstract infrastructure compliant with the CIM specification. Together, they constitute the PIM model, from which different future PSMs shall derive. They both are abstract since their description is devoid of elements relating to their implementation. Moreover, to benefit from the capabilities of the TOSCA standard, we chose to resolutely rely on its normative modeling language to express the PIM model.

The PSM, generated by the third phase, includes (i) the script whose instructions can be interpreted by a dedicated execution environment and (ii) the operational context whose state will be affected by the script's instructions according to the attack scenario. The generated script and context are considered concrete because they make possible to apprehend the attack directly from their implementation artifacts.

To demonstrate the feasibility of our approach, we have transformed the PIM model into a PSM specific to the platform we set up. Our platform provides an infrastructure through OpenTOSCA [27] which is a composite software ecosystem. To enable task automation we also integrated the open source IT automation engine Ansible [28]. Therefore, the execution of the attack script takes place within the built infrastructure.

To illustrate our approach, let us consider an attack where the attacker steals the credentials that a victim uses to connect to a shopping website in order to access his/her account. Throughout this paper, we will refer to this attack as "SnifAttack". To carry out the attack, we will assume scenario steps named and described in Table 1. The three first steps exploit the router's configuration and capabilities. In complement, the two following steps exploit disclosure of sensible information encoded in plain text. Finally, the last step realizes the dreaded event.

Table 1. SnifAttack steps.

| # | Name | Description |
|---|------|-------------|
| 1 | Scanning | The attacker scans its local network gateway to find a listening SSH service. |
| 2 | UseOfDefaults | The attacker uses default credentials to take control of the router. |
| 3 | Sniffing | The attacker has the router do the collecting of all traffic passing through. |
| 4 | Disclosure | The victim sends his/her credentials to log on to the website. |
| 5 | Discovery | The attacker finds out the victim credentials from reading the collected traffic. |
| 6 | Checkmate | The attacker authenticates with the victim's credentials on the shopping website. |

From the point of view of technical networking, the SnifAttack infrastructure relies on 4 major hosts: the attacker's, a router, the victim's PC and the shopping website. Those hosts can reach one another via 3 networks: (i) the one that connects the attacker and the router, (ii) an adjacent one that connects the victim and the router, and (iii) the Internet that connects the router and the remote website. Thus, data sent from one of those networks to a remote one may reach it if the data is forwarded by the router's routing service.

For the script generation corresponding to this scenario, the designer benefits from the guidance provided in the user interface for specifying the attack operating mode requirements. Based on this specification, our approach deduces all the resources involved in carrying out this attack, i.e. the context. For this purpose, we choose Neo4J as the storage system. Fig. 2 is a very small extract of the generated graph from the formal specification of the SnifAttack scenario. The left side of the figure shows that the four infrastructure hosts are connected to their networks and that the routing service is provided by the router. The right side expresses the fact that the scanning functionality is offered by the scanner software installed on the attacker's host device.

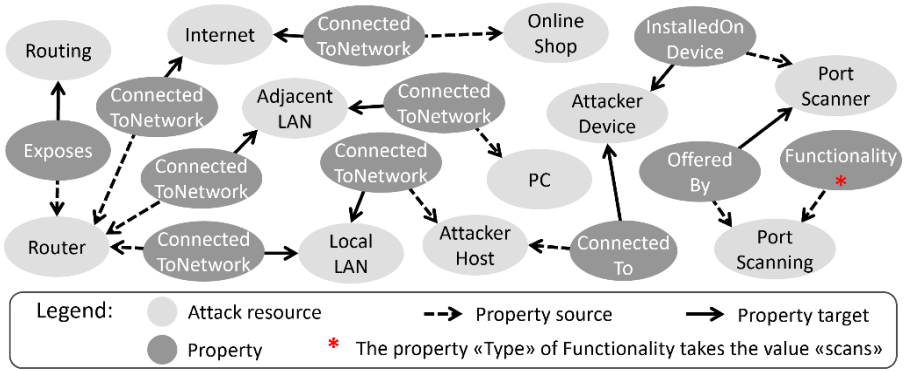

Fig. 2. Excerpt of the SnifAttack knowledge graph

In the PIM level our approach automatically supplies the SnifAttack abstract infrastructure with the topology depicted in Fig. 4 on which the SnifAttack scenario steps will be called according to its abstract script. The latter is also generated automatically by our approach. It is presented as a state diagram (see Fig. 3). Each block of the state diagram represents a step of the attack scenario. It mentions, for example, that the first *step* is triggered when the `scans` action is called on `AttackerHost` (Fig. 3). This action is a *functionality offered by* `PortScanner`. The latter appears in the produced topology deployment diagram depicted in Fig. 4, more specifically in the component `AttackerHost` attached to the `AttackerDevice` storage block.

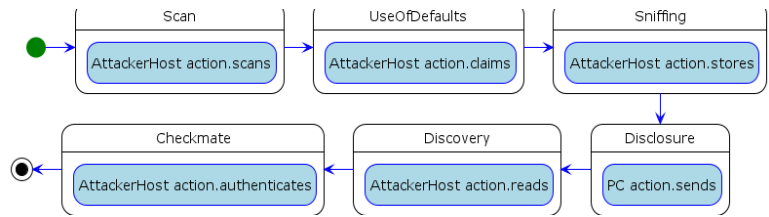

**Fig. 3.** Transitions diagram for SnifAttack generated abstract script workflow

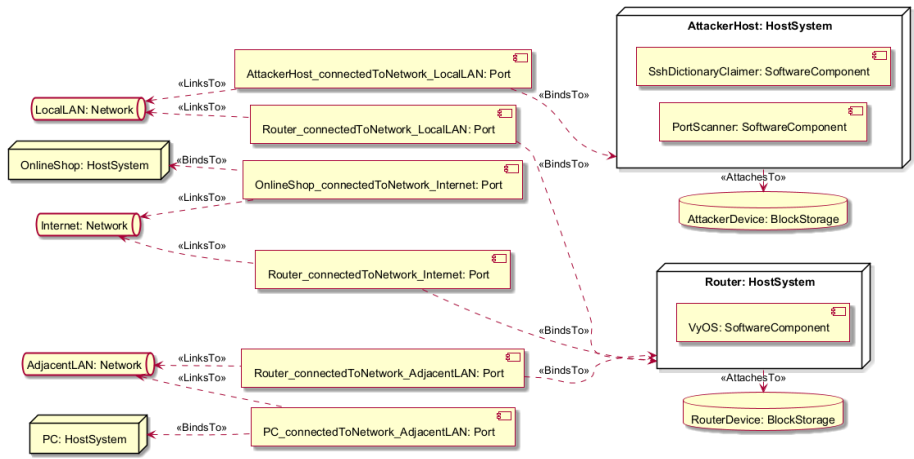

**Fig. 4.** Deployment diagram for SnifAttack topology template

Finally, the two artefacts produced in the PIM level are automatically transformed into specific PSMs in order to be operational on the platform we built. Section 6 supplies an illustrative PSM. Fig. 5 presents the command lines produced in the orchestrator of the platform we built. These command lines correspond to an Ansible execution of the SnifAttack concrete script.

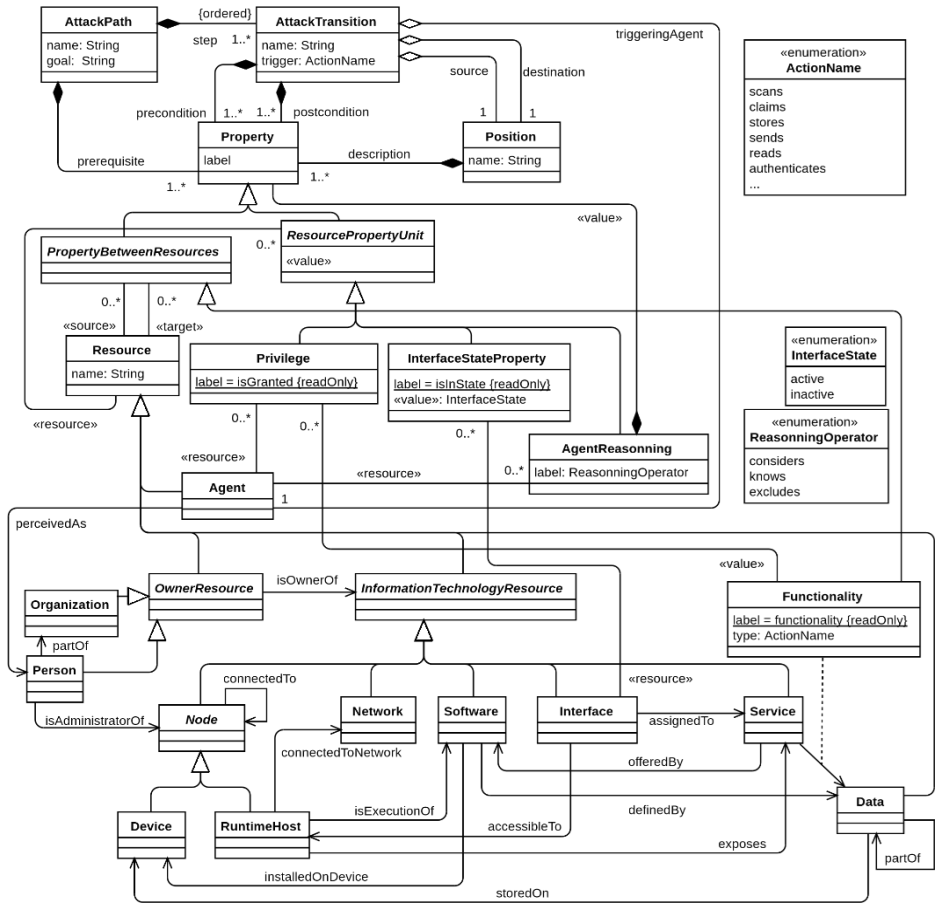

**Fig. 5.** Extract of SnifAttack concrete execution command line output

## 4      Requirement specification

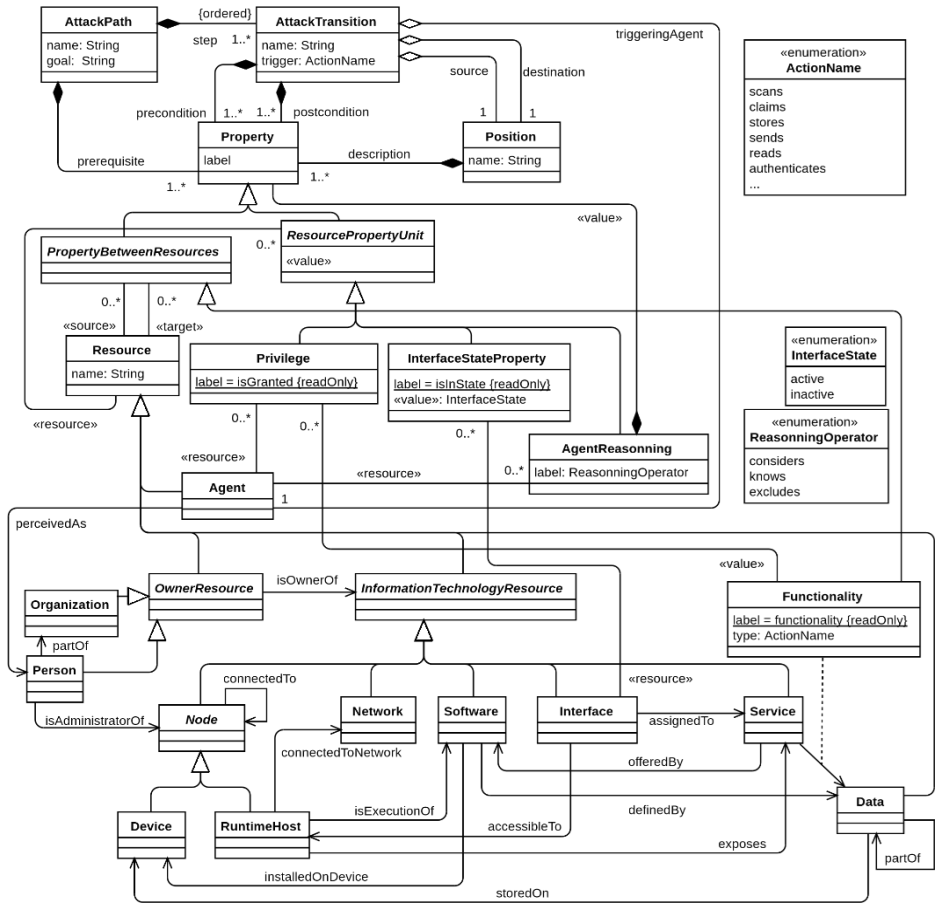

**Fig. 6.** UML class diagram describing our unified attack model

Cybersecurity exercises require above all the availability of attack scripts. The latter, if not available, are usually produced manually from attack scenario specification. An attack scenario specification could be achieved using various models and frameworks proposed by Cyber Threat Intelligence [29] (like Lockheed Martin's Cyber Kill Chain, the MITRE ATT&CK Framework and the Diamond Model) or graph-based attack models provided by academic literature. By automating the script production process, effort can be reduced and errors can be avoided. This automation, although beneficial, remains a challenge or even impossible if it takes as input a specification expressed in any existing model or framework. Indeed, these models and frameworks are syntactically and semantically heterogeneous and they do not share the same knowledge. Moreover, most of them do not offer the possibility to specify the attack context. To overcome these inconvenient issues, we proposed in [3] a unified model that integrates a formalization of the specification of an attack operating mode as well as that of the attack context (i.e the description of all IT resources the attack involves). This model expressed using UML formalism is depicted in Fig. 6. In our model, an attack context is expressed through all the resources that could be involved in an attack. They are generalized in the class *Resource* which includes among others the resources *Runtime-host*, *Software* and *Network*. Relationships between resource classes are also described in our model. Labels used for their description are part of the controlled vocabulary we suggested for the formalization of the attack operating mode description. The latter is expressed through the class *AttackPath*. Like in the state-enumeration models, an attack path looks like a state-transition model where states correspond to the context states and where transitions describe changes in the context state. To express postconditions and preconditions of transitions two kinds of resources properties are used: those characterizing resources and those describing connection between them.

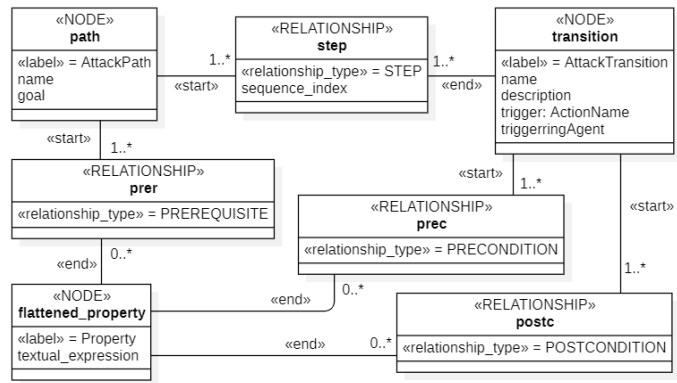

**Fig. 7.** Part of the knowledge graph structure describing an attack operating mode

To guide a user through the specification task, an interface is proposed. It provides him/her several patterns that can be used to specify the facts that constitute the operating mode. The formalization of the attack operating mode makes it possible to automatically generate the specification of the attack context, which includes all the resources involved in the attack, mainly those targeted by the attack and those mobilized by the attacker to achieve his/her goal. The two specifications are stored together as a knowledge graph in a property graph database whose structures are respectively described in Fig. 7 and Fig. 8. In case of the formalized SnifAttack, 38 resources were derived, making the global knowledge graph reach 182 nodes and 1482 relationships. Fig. 2 is a very small extract of the SnifAttack knowledge graph.

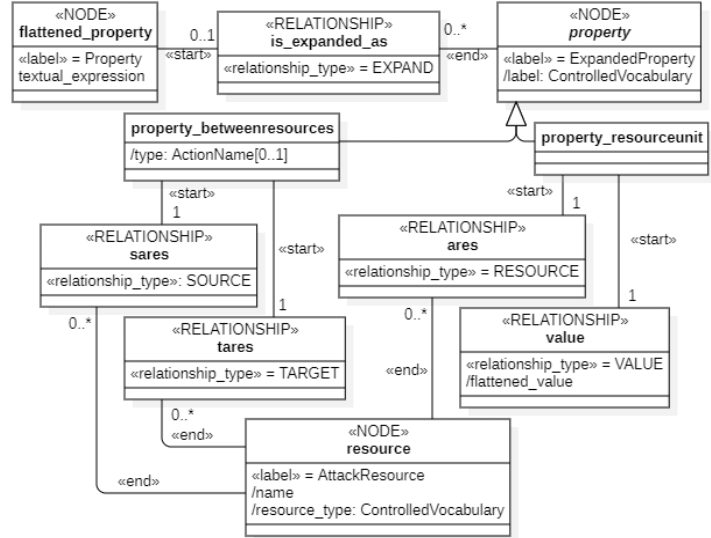

**Fig. 8.** Part of the knowledge graph structure describing the attack context

## 5    PIM generation

For the expression of the PIM model (abstract script and abstract infrastructure) we chose to use the TOSCA language.

TOSCA is an OASIS open standard that defines the interoperable description of services and applications hosted on the cloud and elsewhere [30]. Its use strengthens the application (i) by modeling its needs at different development phases; and (ii) by allowing its reuse on different platforms (interoperability between cloud providers). Modeling in TOSCA was originally crafted with the microservice architecture point of view. Thus, the application model is formulated as a service template which comprises the description of the infrastructure topology and its related workflows. The workflows are responsible for changing the topology states with defined series of activities.

In our case, the abstract infrastructure corresponds to the topology template, in the initial context state, while the abstract script corresponds to a dedicated workflow which is able to change the state of the topology template.

Therefore, we defined transformation rules to automatically generate both the topology template and the related workflow from the knowledge base.

Let us note that we have encoded the service template description in YAML by using the TOSCA Simple Profile version 1.3 [31], thus benefiting from its normative types.

Therefore, our PIM automatic generation process encompasses the steps presented in Fig. 9. First, the template initialization loads invariants from Lst 2. The latter includes complementary profiling for the generation of an attack workflow. Indeed, TOSCA language was intended for modeling generic services and operations. This constitutes a limitation for expressing attacks that may require vulnerability exploit and/or misuse of services. Thus, we made a minimal enrichment which (i) describes the operations to carry out the attack in a dedicated interface type, and (ii) enables their call on hosts in the infrastructure. This enrichment consists in adding the concepts `AttackTransitions` and `HostSystem` in the service template. The next two steps of our PIM automatic generation process focus on the automatic transformation from the CIM to the PIM. The two paragraphs of this section provide detailed information about them.

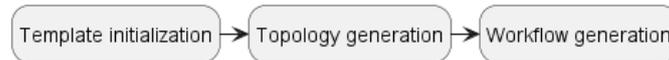

**Fig. 9.** PIM generation steps

**Lst 2.** YAML primitive content for PIM

```
tosca_definitions_version: tosca_simple_yaml_1_3
interface_types:
   AttackTransitions:
      derived_from: tosca.interfaces.Root
node_types:
   HostSystem:
      derived_from: Compute
      interfaces:
         action:
            type: AttackTransitions
```

To realize the automatic transformation from the CIM to the PIM, we designed a set of twelve rules, each one is a couple of two patterns. The first one, called *trigger pattern*, enables the rule to match instances in the knowledge base encoding the CIM. It also allows the extraction of data from their matching instances for applying the rule. The second one, called *effect pattern*, fills the service template encoded in YAML. Due to lack of space, only few of them are presented in this paper. The two following paragraphs will present generation from an instance of a CIM model of respectively the topology infrastructure and the workflow.

**Topology generation.**
The topology relates to the context in the state before any attack step had been carried. Thus, for its generation trigger patterns are applied to context elements supplied by the first position.

Based on types, some CIM resources are directly translated into PIM node templates. For instance, Fig. 10 presents rules implementing the host-centric topology design intended with the profiling. Rule 1 turns each *runtime-host* into a `HostSystem`, and rule 2 turns each *network* into a `Network`. Besides those simple resource conversions, topology generation also translates more complex expressions, composed with properties. For instance, rule 3 (Fig. 10) turns each network connection into a `Port`.

| Rule | Trigger pattern | Effect pattern |
|---|---|---|
| 1 | n:resource, name, resource_type='RuntimeHost' | ```
topology_template:
  node_templates:
    {n.name}:
      type: HostSystem
``` |
| 2 | n:resource, name, resource_type='Network' | ```
topology_template:
  node_templates:
    {n.name}:
      type: Network
``` |
| 3 | SOURCE → n1:resource, name, resource_type='RuntimeHost'; :property_betweenresources, label='connectedToNetwork'; TARGET → n2:resource, name, resource_type='Network' | ```
topology_template:
  node_templates:
    {n1.name}_connectedToNetwork_{n2.name}:
      type: Port
      requirements:
        - link: {n2.name}
        - binding: {n1.name}
``` |

**Fig. 10.** Sample rules for the topology generation

We applied the rules on our SnifAttack formalized attack, Lst 3 represents an excerpt of the topology generated by rules from above. The `AttackerHost` and `LocalLAN` declarations are the results of rules 1 and 2, while `AttackerHost_connectect-edToNetwork_LocalLAN` which refers to them is the result of rule 3.

**Lst 3.** SnifAttack topology networking excerpt

```
...
topology_template:
  node_templates:
    AttackerHost:
       type: HostSystem
    LocalLAN:
       type: Network
    AttackerHost_connectedToNetwork_LocalLAN:
       type: Port
       requirements:
          - link: LocalLAN
          - binding: AttackerHost
```

**Workflow generation.**
As mentioned above, the abstract script generation corresponds to the attack operating mode description as a dedicated workflow.

First, `AttackTransitions` is profiled with operations matching the functionalities triggered in scenario steps, which are formalized by transitions. For that purpose, rule 4 (Fig. 11) hands those operations, with their description, to the interface type.



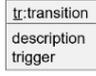

**Fig. 11.** Workflow generation rules

Then, the attack path information, that describes the carried-out attack scenario in CIM, is appended to the dedicated `AbstractScript` workflow. Therefore, its goal describes the workflow, rule 5 (Fig. 11) conveys it in the description field. The steps mentioned in the scenario are also translated to the script's workflow, and their triggering by agents is automated in the workflow by operation calls, intended to be operated by a TOSCA orchestration process. Rule 6 (Fig. 11) implements the nested correspondence. For sequencing within the workflow according to the imperative TOSCA way, each step in the workflow specifies the following one, rule 7 (Fig. 11).

**Lst 4.** Effect pattern for filling steps with the target

```
topology_template:
  workflows:
    AbstractScript:
      steps:
        {tr.name}:
          target: {sys.name}
```

At this point, each step of the workflow still lacks the `HostSystem` target to call its operations. Finding targets enable to leverage Lst 4 for completing the workflow. Although this information does not directly appears in the scenario database, it can be inferred by reasoning, empowered by the conceptual model [3], on its knowledge graph. Thus, we formulated hypotheses on CIM which, when their assumption matches in the knowledge graph, reveal the host on which to target the attack step operation.

In the following, we present two of them, respectively called `iao` and `ig`.



*Hypothesis* `iao`.

CIM formulation: If *the agent `<agt>` triggers the transition `<tr>` with the functionality `<func>` offered by a software installed on a host `<sys>` where the agent is perceived as administrator*, then the step's trigger operation is called this *host `<sys>`*.

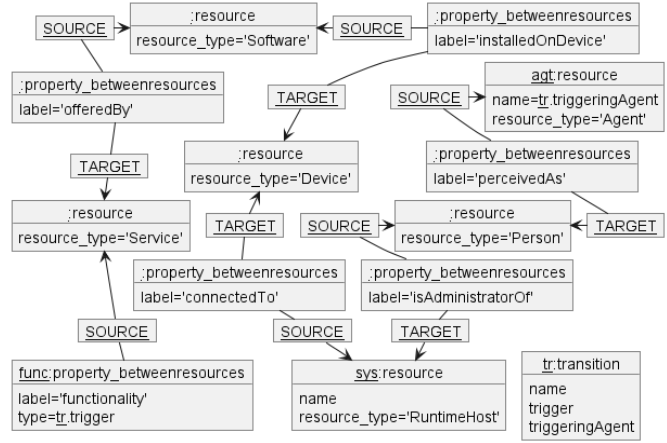

**Fig. 12.** Trigger pattern matching iao assumption

In order to seek out matching entities in the knowledge graph, we crafted the corresponding query whose *trigger pattern* is depicted Fig. 12.

In case of SnifAttack, `iao` successfully infers targets for the first two path steps: `Scan` and `UseOfDefault` (Lst 5).

To take into account that agents can also leverage remote functionalities on hosts they compromised, we had to slightly extend the hypothesis assumption. In case of the SnifAttack, the extended version infers targets of two additional path steps: `Sniffing` and `Disclosure` (Lst 5). In the attack narrative, it corresponds to capabilities the attacker acquires by compromising the router.

*Hypothesis* `ig`.

CIM formulation: If *the agent `<agt>` is granted the functionality `<func>` that he/she triggers in the transition `<tr>` via an interface accessible from a host `<sys>` where the agent is perceived as administrator*, then the step's trigger operation is called this *host `<sys>`*.

The *trigger pattern* corresponding to the `ig` query is depicted Fig. 13.

In case of SnifAttack, it finds targets for the two remaining steps: `Discovery` and `Checkmate` (Lst 5).



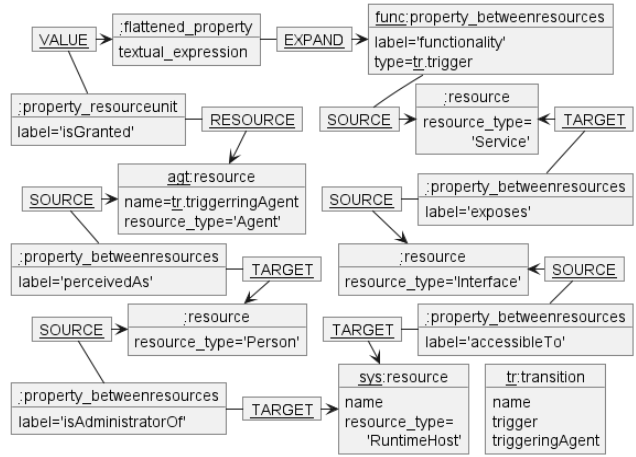

**Fig. 13.** Trigger pattern matching ig assumption

**Lst 5.** Excerpt of the SnifAttack generated workflow

```
...
topology_template:
   ...
   workflows:
     AbstractScript:
       steps:
         Scan:
           activities:
             - call_operation: action.scans
           on_success: [ UseOfDefaults ]
           target: AttackerHost
         UseOfDefaults:
           activities:
             - call_operation: action.claims
           on_success: [ Sniffing ]
           target: AttackerHost
         Sniffing: ...
           target: AttackerHost
         Disclosure: ...
           target: PC
         Discovery: ...
           target: AttackerHost
         Checkmate: ...
           target: AttackerHost
```

Once a target is declared for every step in the `AbstractScript` workflow, the resulting PIM is a complete TOSCA service template description. Therefore, it may be leveraged by any stakeholder through the standard.

For the SnifAttack example, we used the TOSCA Toolbox [32, 33] to empirically check the generated service template. After passing the standard semantic and syntactic check, the toolbox automatically generated the UML diagrams Fig. 4 and Fig. 3.



## 6     PSM generation

By adopting MDA principles for script generation, we defined three layers of abstraction that separate the business and application logic from the platform technology used for the emulation of an attack, according to a provided concrete script ans concrete attack infrastructure. The emulation consists of reproducing the behavior of both the attack operating mode and the context components. However, this concrete script and concrete infrastructure to be produced in the PSM generation phase would need to be deduced from their abstract expressions produced in the previous phase using the chosen pecific platform. Lacking a cyber-range type of platform, we opted for a virtual laboratory on a server where attacks can be implemented and deployed. Our platform is mainly composed of two components: the OpenTOSCA ecosystem [27, 34], for providing the infrastructure, and Ansible [28], for automated orchestration.

OpenTOSCA offers a chained integration of specialized tools. Its primary user entrance is a web interface which distinguishes content across tabs. For the emulation we are aiming for, the two relevant tabs are 'repository' and 'applications', which respectively display on-going and ready-to-deploy projects. Winery is the component handling on-going projects in the ecosystem. It contains a modeling tool for drawing most of the concrete infrastructure as a topology sketch, though it falls short on network configuration. Once the project is completed, the user can install it through the primary web interface, making it usable by the ready-to-instantiate applications.

Ansible is an open-source IT automation engine. It operates at server level. Its projects are specified as tasks defined in playbooks. At runtime, the configuration of the defined tasks is derived from the provided inventory.

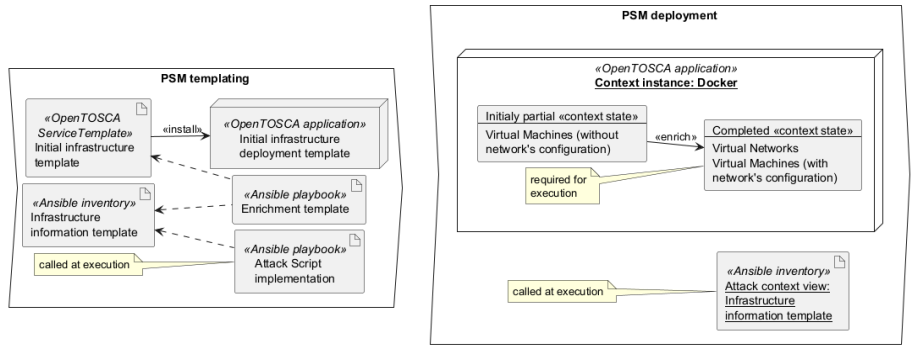

**Fig. 14.** Summarized platform specific PSM generation process

The PSM generation process encompasses two phases: PSM templating and PSM deployment (Fig. 14). The first phase is dedicated to the generation of four deployment templates: the OpenTOSCA application for the initiation of the concrete infrastructure instance, the infrastructure information template, the automatic enrichment script for completing the networking configuration in the instance, and the attack script for automating the attack execution. The two scripts correspond to Ansible playbooks. Therefore, they can share the inventory that configures their tasks. The infrastructure



information template (i.e. Ansible inventory template) will contain the arborescence of references to hosts.

The second phase deploys the attack thanks to the generated templates. In other words, the automated deployment implements the concrete attack instantiation, enabling the attack script to run in the concrete attack infrastructure.

In the case of the SnifAttack example, the Ansible inventory template is depicted in Lst 6. Lst 7 supplies the concrete script related to the SnifAttack example. Fig. 5 shows an execution of the concrete script of the SnifAttack example. This execution is done on the concrete infrastructure 2866 mentioned in the first line corresponding to the command calling the attack script playbook.

**Lst 6.** Extract of the Ansible inventory template for SnifAttack example

```
all:
   children:
      Agent:
         children:
            Attacker:
            ActingVictim:
      Attacker:
         hosts:
            AttackerHost:
      ActingVictim:
         hosts:
            PC:
```

**Lst 7.** SnifAttack's concrete attack script suited for inventory Lst 6

```
---
- name: Scan (Attacker scans) - The attacker scans its local network gateway, …
  hosts: Attacker
  roles:
     - AttackTransition_Scan
- name: UseOfDefaults (Attacker claims) - The attacker uses default credential…
  hosts: Attacker
  roles:
     - AttackTransition_UseOfDefaults
- name: Sniffing (Attacker stores) - The attacker has the router do the collec…
  hosts: Attacker
  roles:
     - AttackTransition_Sniffing
- name: Disclosure (ActingVictim sends) - The victim sends his/her credentials…
  hosts: ActingVictim
  roles:
     - AttackTransition_Disclosure
- name: Discovery (Attacker reads) - The attacker finds out the victim`s crede…
  hosts: Attacker
  roles:
     - AttackTransition_Discovery
- name: Checkmate (Attacker authenticates) - The attacker authenticates with t…
  hosts: Attacker
  roles:
     - AttackTransition_Checkmate
```

## 7   Conclusion

In this paper we addressed the challenging issue of automating the generation of attack script and attack context from informal description of scenarios. A model-driven approach is proposed. It separates business and application design from underlying



platform technology by defining abstraction levels of the software to be developed. Thanks to our adoption of the MDA method and the TOSCA standard, our approach promotes intensive reuse at all abstraction levels and multi-platform implementation. We also proposed our rule-based transformation mechanisms for automating the generation of abstract attack scripts and infrastructure from formalized requirements. The approach has been implemented and tested for the generation of script corresponding to a complex attack.

We are currently working on the user-friendly design of the application to assist security experts in the attack scenario formalization. The present research work contributes to the reduction of effort in script production. The implementation of an evaluation process is in progress. Future work will go towards reuse.

**Acknowledgments.** The authors thank Prof. Françoise SAILHAN for the advice provided during this research work.